\documentclass[prd,aps,twocolumn,amsmath,amssymb,nofootinbib,preprintnumbers]
{revtex4}

\usepackage{graphicx}
\usepackage{dcolumn}
\usepackage{bm}
\usepackage{amsmath}
\usepackage{amsfonts}
\usepackage{euscript,bbm}
\usepackage{ifthen}
\usepackage{psfrag}

\def\ls{\mathrel{\lower4pt\vbox{\lineskip=0pt\baselineskip=0pt
           \hbox{$<$}\hbox{$\sim$}}}}
\def\gs{\mathrel{\lower4pt\vbox{\lineskip=0pt\baselineskip=0pt
           \hbox{$>$}\hbox{$\sim$}}}}
\def\drawbox#1#2{\hrule height#2pt

\hbox{\vrule width#2pt height#1pt \kern#1pt
              \vrule width#2pt}
              \hrule height#2pt}

\def\Asym#1#2{\vcenter{\vbox{\drawbox{#1}{#2}
              \kern-#2pt       
              \drawbox{#1}{#2}}}}

\newcommand{\be}{\begin{equation}}
\newcommand{\ee}{\end{equation}}
\newcommand{\bea}{\begin{eqnarray}}
\newcommand{\eea}{\end{eqnarray}}

\newcommand{\neu}[1]{\ensuremath{\tilde{\chi}_{#1}^0}}
\newcommand{\chp}[1]{\ensuremath{\tilde{\chi}_{#1}^+}}
\newcommand{\chm}[1]{\ensuremath{\tilde{\chi}_{#1}^-}}

\newcommand{\chpm}[1]{\ensuremath{\tilde{\chi}_{#1}^{\pm}}}

\newcommand{\met} {{E\!\!\!\!/_{\rm T}}}
\newcommand{\pT} {{p_{\rm T}}}
\newcommand{\ET} {{E_{\rm T}}}

\newcommand{ \pgs }    {{\tt PGS4}}

\newcommand{ \madgraph }    {{\tt Madgraph}}

\begin{document}

%
\title{Vector Boson Fusion Processes as a Probe of Supersymmetric Electroweak Sectors at the LHC}

\author{Bhaskar Dutta$^{1}$}
\author{Alfredo Gurrola$^{2}$}
\author{Will Johns$^{2}$}
\author{Teruki Kamon$^{1,3}$}
\author{Paul Sheldon$^{2}$}
\author{Kuver Sinha$^{1}$}

\affiliation{$^{1}$~Mitchell Institute for Fundamental Physics and Astronomy, \\
Department of Physics, Texas A\&M University, College Station, TX 77843-4242, USA \\
$^{2}$~Department of Physics and Astronomy, Vanderbilt University, Nashville, TN, 37235 \\
$^{3}$~Department of Physics, Kyungpook National University, Daegu 702-701, South Korea
}

\begin{abstract}

Vector boson fusion (VBF) processes offer a promising avenue to study the non-colored sectors of supersymmetric extensions of the Standard Model at the LHC. A feasibility study for searching for the chargino/neutralino system in the $R-$parity conserving Minimal Supersymmetric Standard Model is presented. The high $\ET$ forward jets in opposite hemispheres are utilized to trigger VBF events, so that the production of the lightest chargino $\chpm{1}$ and the second lightest neutralino $\neu{2}$ can be probed without a bias by experimental triggers. Kinematic requirements are developed to search for signals of these supersymmetric states above Standard Model backgrounds in both $\tau$ and light lepton ($e$ and $\mu$) final states at $\sqrt{s} = 8$ TeV. 


\end{abstract}
MIFPA-12-35
\maketitle


\section{Introduction}

The LHC has put bounds on the masses of the gluino ($\tilde{g}$) and first two generation squarks ($\tilde{q}$). For comparable masses, they are excluded up to about $1500$ GeV at $95\%$ CL at $\sqrt{s}= 7$ TeV with $4.7$ fb$^{-1}$ of integrated luminosity \cite{:2012rz, Aad:2012hm, :2012mfa}. Limits on the masses of the third generation colored superpartners have also been under serious theoretical scrutiny recently \cite{Dutta:2012kx, Plehn:2011tg}, and bounds in certain decay modes 
%
have been obtained from the LHC \cite{ATLASStop}. 

On the other hand, the bounds on charginos and neutralinos are less constrained, since, expectedly, these particles suffer from smaller electroweak (EW) production in a hadron collider. The chargino/neutralino system plays a crucial role in the dark matter connection of supersymmetric models. The lightest neutralino $\neu{1}$, as the lightest supersymmetric particle (LSP), is the canonical dark matter candidate in the $R$-parity conserving minimal supersymmetric extension of the Standard Model (MSSM). Various schemes of supersymmetry breaking and mediation predict $m_{\neu{2}}$ (where $\neu{2}$ is the second lightest neutralino) and $m_{\tilde{\chi}^{\pm}_{1}}$ (where $\tilde{\chi}^{\pm}_{1}$ is the lightest chargino) to be larger by a factor of $1-3$ than $m_{\neu{1}}$, under broad assumptions \cite{Choi:2007ka}. For Higgsino-like $\neu{1}$, the degeneracy is much stronger \cite{Allahverdi:2012wb}. This implies that if $m_{\neu{1}}$ is in the mass range of $\mathcal{O}(100)$ GeV, the $\neu{2}, \chpm{1}$ system may be expected to be in the $100-300$ GeV range, which is within the range of detection at the LHC given suitable search strategies.



 From the perspective of a hadron collider, where EW production is small, a classic strategy to study the chargino/neutralino system is to detect the neutralinos in cascade decays of gluinos and squarks. For example, reconstructing a decay chain like $\tilde{g} \, \rightarrow \, \tilde{q} \, \rightarrow \, \neu{2} \, \rightarrow \, \tilde{\tau}_1 \, \rightarrow \, \neu{1}$ using endpoint methods \cite{Arnowitt:2008bz, Hinchliffe:1996iu} leads to  mass measurements of $\neu{2}$, $\tilde\tau_1$, $\tilde q$ and $\tilde g$,  where $\tilde\tau_1$ is the lighter stau mass \cite{Dutta:2011kp}. However, in a scenario where colored objects are heavy and the production cross-section is limited, one has to use different techniques to probe the EW sector. 

\begin{figure}[!htp]
\centering
\includegraphics[width=2.0in]{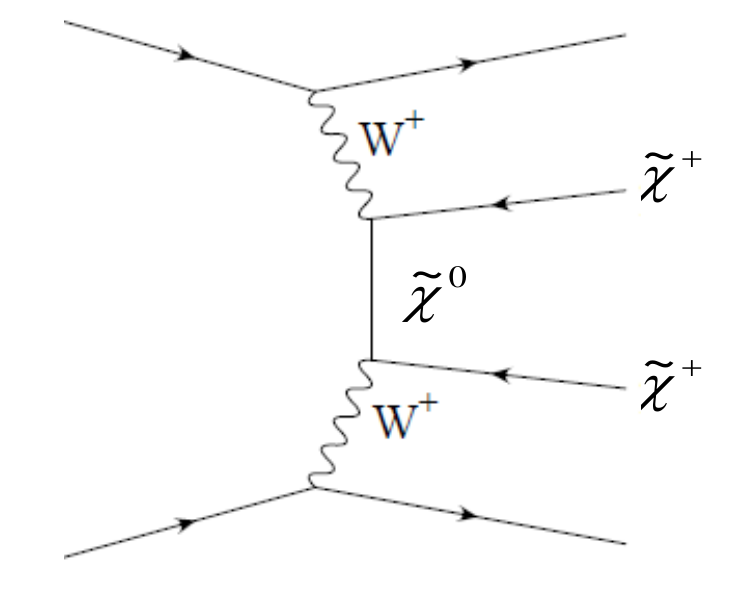}
\caption{Feynman diagram of production of charginos pairs by VBF processes.}
\label{VBFCharginoDiagram}
\end{figure}

The purpose of this paper is to show that the chargino/neutralino system can be studied in events through vector boson fusion (VBF) processes \cite{Cahn:1983ip, Bjorken:1992er}.
%
  VBF processes have been suggested for Higgs searches \cite{Rainwater:1998kj} and supersymmetric searches, in the context of slepton and gaugino productions at $14$ TeV LHC \cite{Choudhury:2003hq,cho,datta}. VBF production is characterized by the presence of two jets with large dijet invariant mass in the forward region in opposite hemispheres. A sample diagram of chargino production from VBF processes is shown in Figure \ref{VBFCharginoDiagram}.

Direct searches for $\neu{2}$ and $\chpm{1}$  at ATLAS \cite{ATLASStrilepton, ATLAS:2012ab} and CMS \cite{Chatrchyan:2012mea} mainly involve the search for signal in events with $3$ leptons and $\met$: $\tilde{\chi}^{\pm}_1 \, \neu{2} \,\, \rightarrow \,\, l l l \nu \neu{1} \neu{1}$ \footnote{For example, ATLAS bounds in the trilepton $+ \met$ searches at $\sqrt{s} = 7$ TeV, with $4.7$ fb$^{-1}$ of integrated luminosity rule out charginos with mass up to about $500$ GeV for $m_{\neu{1}} \, < \, 230$ GeV and large branching ratios of $\chpm{1}$ and $\neu{2}$ to a slepton. In anomaly-mediated supersymmetry breaking scenarios, ATLAS has placed a bound on the chargino mass of $90$ GeV, for $m_{3/2} < 32$ TeV, $m_0 < 1.5$ TeV, ${\rm tan} \beta = 5$ and $\mu > 0$.}. This can proceed through $\chpm{1}$ decaying through $(\tilde{\nu}l)$, $(\tilde{l}_1 \nu)$, or $(W^{\pm (*)}\neu{1})$, and $\neu{2}$ decaying to $(\tilde{l}_1 l)$, $(\tilde{\nu}\nu)$, or $(Z^{(*)}\neu{1})$. 

%
Several features of existing searches at the LHC may be highlighted for better contrast with the present study:

$(a)$ Searches target $\neu{2}$ and $\chpm{1}$ produced via Drell-Yan production, i.e., without requiring the presence of forward/backward jets.

$(b)$ The bounds assume $m_{\tilde{\chi}^{\pm}_1} \, \sim \, m_{\neu{2}}$ and an enhanced branching ratio to trilepton final state provided by $m_{\tilde{l}_1} \sim (m_{\tilde{\chi}^{\pm}_1} + m_{\neu{1}})/2$ (where $\tilde{l}_1$ is the lighter slepton, and $l$ denotes either $e$ or $\mu$). The slepton mass $m_{\tilde{l}_1}$ is carefully placed such that the mass splitting is large and thus the leptons have relatively high $p_{T}$ to be free from trigger bias.

$(c)$ Searches resulting in $\tau$ final states do not exist due to the larger $\tau$ misidentification rates which make it difficult to both manage the level of backgrounds and maintain low enough $p_{T}$ thresholds for triggering.

In this paper, final states with  $2 \tau \, + \, \met$ as well as $2 l \, + \, \met$ (along with two high $\pT$ forward/backward jets) arising from $\chpm{1}$ and $\neu{2}$ produced by VBF processes  are studied. This offers certain advantages:

$(i)$ With increasing instantaneous luminosity at the LHC, both ATLAS and CMS experiments are raising their $\pT$ thresholds for triggering any object. This motivates us to probe signals for supersymmetry in VBF processes at $\sqrt{s} = 8$ TeV where production of superpartners in VBF processes are free from trigger bias.

$(ii)$ VBF production allows the investigation of final states with $\tau$. $\tilde\tau_1$  is typically lighter than $\tilde\mu_1$ and $\tilde e_1$ for large $\tan\beta$. A light $\tilde{\tau}_1$ with small mass splitting is favored in coannihilation processes \cite{Griest:1990kh} that set the relic density to correct values, in the case of Bino dark matter. Light $\tilde{\tau}_1$ is also motivated in the context of the MSSM by the enhancement of the $h \, \rightarrow \, \gamma \gamma$ channel \cite{Carena:2011aa}. These facts stress the importance of searches in $\tau$ final states with low $\pT$ and large backgrounds, for which production by VBF processes is more suited since the VBF signature allows for the reduction of the backgrounds to manageable levels.

$(iii)$ For the leptonic final state, a search based on VBF processes can be complementary or better than the existing LHC searches based on Drell-Yan production, since it is not constrained by the trigger bias. It is also interesting to note that the Drell-Yan production cross-section falls faster than the VBF production cross-section with increasing mass \cite{Choudhury:2003hq}. 





The structure of the paper is as follows. In Section \ref{searchstrategy} an outline of the search strategy is given, followed by results in Section \ref{results}. Conclusions are given in Section \ref{conclusion}.



\section{Search Strategy} \label{searchstrategy}





The $\neu{2}$ and $\chpm{1}$ are produced by VBF processes and then decay into  the lighter slepton states ($\tilde\tau_1$,  $\tilde\mu_1$ and  $\tilde e_1$) by the decay processes $\tilde{\chi}^{\pm}_1 \, \rightarrow \, \tilde{\tau}_1 \nu \, \rightarrow \, \neu{1} \tau \nu$, $\tilde{\chi}^{\pm}_1 \, \rightarrow \, \tilde{ l} \nu \, \rightarrow \, \neu{1} l \nu$, and similarly for $\neu{2}$ via $\neu{2}\, \rightarrow \, \tilde{\tau}_1 \tau \, \rightarrow \, \neu{1} \tau \tau$ and $\neu{2}\, \rightarrow \, \tilde{l}_1 l \, \rightarrow \, \neu{1} ll$. 


%
Depending on the branching ratio of $\neu{2}$ and $\chpm{1}$ to $\tilde l$ and $\tilde{\tau}_1$  the results presented here can be used in feasibility studies of detecting different  models with smaller chargino/neutralino masses around the EW scale at the LHC.

A benchmark point is first defined and the following processes are investigated:
%
%
\be
pp \rightarrow \chpm{1} \, \chpm{1} \, jj , \,\,\, \chp{1} \, \chm{1} \, jj , \,\,\, \chpm{1} \, \neu{2} jj ,\,\,\, \neu{2} \, \neu{2} jj \,\,.
\ee
The benchmark point is $m_{\tilde{\chi}^{\pm}_1} \, \sim \, m_{\tilde{\chi}^0_2} \, = 181$ GeV, $m_{\tilde{\tau}_1} = 130$ GeV, and $m_{\tilde{\chi}^0_1} = 100$ GeV. The $\chpm{1}$ and \neu{2} are mainly Wino, while $\neu{1}$ is mainly Bino.



%
%
%


The search strategy is based on two steps: first, use the unique features of VBF processes to reduce background $V+$jets events (where $V$ is either $W$ or $Z$), and second, use decay properties of the centrally produced supersymmetric particle to reduce non-supersymmetric channels that are also produced by VBF processes.

$(1)$ VBF processes: VBF processes are characterised by the production of two energetic jets produced in the forward-backward regions with large $|\Delta \eta|$. Requirements on $|\Delta \eta|$ as well as $M_{j_1j_2}$, where $M_{j_1j_2}$ is the dijet invariant mass, are effective in reducing $V+$jets and $t \overline{t}$ background events, where the jets are less energetic and more central. Moreover, the leading jet is expected to have high $p_{T}$ since the incoming partons need significant momentum to create a pair of heavy supersymmetric states. This motivates a $p_T$ cut on the leading jet to reduce background.

$(2)$ Decay of $\chpm{1}$ and $\neu{2}$: Having reduced the background using the above selections, the remnant background consists of other particles that have been produced centrally, also by VBF processes. For $R-$parity conserving models, the decay of the charginos and neutralino proceeds as
\bea
\chpm{1} \, \rightarrow \, \tilde{\tau}^{\pm}_1 \nu \, \rightarrow \, \tau^{\pm} \neu{1} \nu \nonumber \\
\neu{2}  \, \rightarrow \, \tilde{\tau}^{\pm}_1 \tau^{\mp} \, \rightarrow \, \tau^{\pm} \tau^{\mp} \neu{1} \,\,.
\eea
The $\neu{1}$ LSP escapes the detector as the dark matter candidate, and gives rise to $\met$. 

The production of  $VV$ (where $V$ may be wither $W$ or $Z$) by VBF processes mimics the signal when the bosons decay leptonically. A $\met$ cut is effective in reducing this background. 
Moreover, requiring multiple $\tau$'s in the event further reduces background. Results will be presented for requiring same-sign and oppositely-signed $\tau$ pairs, as well as an inclusive study. 
%
%
Although  $m_{\tilde{\chi}^{\pm}_1} \, \sim \, m_{\tilde{\chi}^0_2}$ is chosen as an example, the methods described in this paper are applicable in detecting $\neu{2}$ and $\chpm{1}$ separately.

After presenting results in the $2 \, \tau$ event selection, further results will be presented for $2 \, l$ event selection, corresponding to 
\be
\neu{2} \, \rightarrow \, \tilde{l}^{\pm}_1 l^{\mp} \, \rightarrow \,  l^{\pm} l^{\mp} \neu{1} \,\,.
\ee

\section{Results} \label{results}

In this section, the main results are presented. 
Signal and background samples are generated with \madgraph \,\, \cite{Alwall:2011uj} followed by detector simulation using \pgs \, \cite{pgs}.

For the VBF selections, motivated in the search strategy in Section \ref{searchstrategy}, we accept jets with $\pT \geq 50$ GeV in $|\eta| \leq 5$, and require a presence of two jets ($j_1$, $j_2$) satisfying: 

$(i)$ $\pT(j_1)  \geq 75$ GeV;

$(ii)$ $|\Delta \eta (j_1, j_2)| > 4.2$;

$(iii)$ $\eta_{j_1} \eta_{j_2} < 0$;

$(iv)$ $M_{j_1j_2} > 650$ GeV.

We note that the signal acceptance with this selection  is less sensitive to effects on the signal acceptance due to initial/final state radiation, pileup, and fluctuations in jet fragmentation.

\begin{figure}[!htp]
\centering
\includegraphics[width=3.5in]{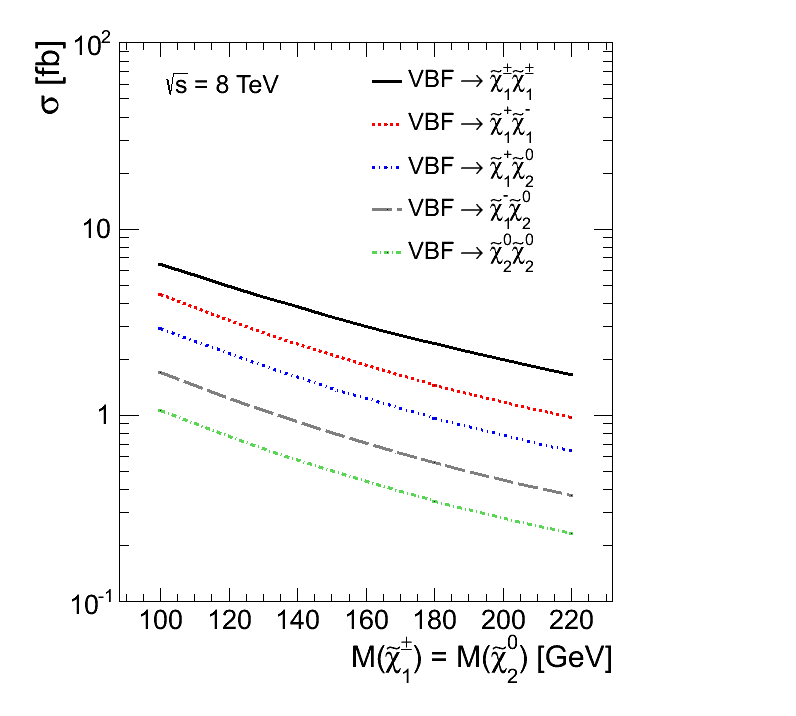}
\caption{VBF production cross-section at $\sqrt{s} = 8$ TeV as a function of mass for various channels after imposing $|\Delta \eta| \, > \, 4.2$ using Madgraph 4.}
\label{VBFXsection}
\end{figure}

The production cross-sections at $\sqrt{s} = 8$ TeV for $\chpm{1} \chpm{1}, \chp{1} \chm{1}, \chpm{1} \neu{2},$ and $\neu{2} \, \neu{2}$ as a function of mass after imposing just $|\Delta \eta| \, > \, 4.2$ are displayed in Figure \ref{VBFXsection}. 
For the $\chpm{1} \chpm{1}$ mode, the production cross-section is $\sim 7$ fb for $m_{\tilde{\chi}^{\pm}_1} \, \sim \, 100$ GeV, while for oppositely signed charginos, the corresponding cross-section is $\sim 5$ fb. At the benchmark value of $m_{\tilde{\chi}^{\pm}_1} \, = \, 181$ GeV, the corresponding values are $2.5$ fb and $1.5$ fb respectively. We have used Madgraph 4 for our calculation.


The appreciable production cross-sections for $\chpm{1} \chpm{1}$ and $\chpm{1} \neu{2}$ mean that same-sign $\tau$'s are significantly present in the final state. As will be seen, the same-sign $\tau$ selection leads to considerable reduction of background.


\begin{figure}[!htp]
\centering
\includegraphics[width=3.5in]{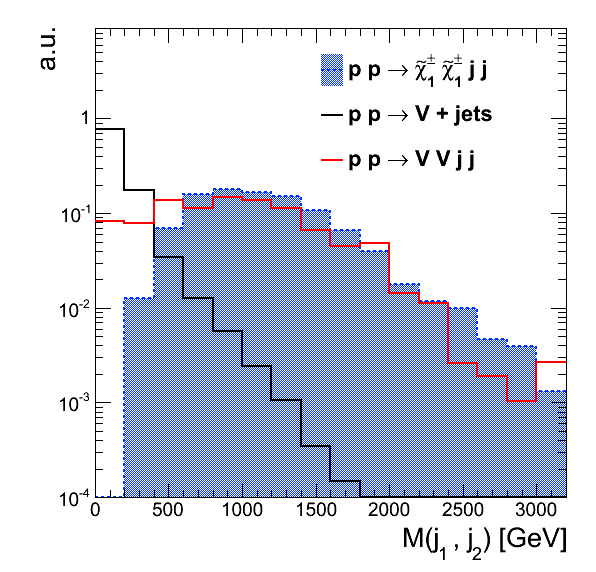}
\caption{$M_{j_1j_2}$ distribution normalized to arbitrary units for $\chpm{1} \chpm{1}$ pair production by VBF processes, $V+$jets background, and $VV$ background produced by VBF processes.}
\label{VBFdistributionMj1j2}
\end{figure}

\begin{figure}[!htp]
\centering
\includegraphics[width=3.5in]{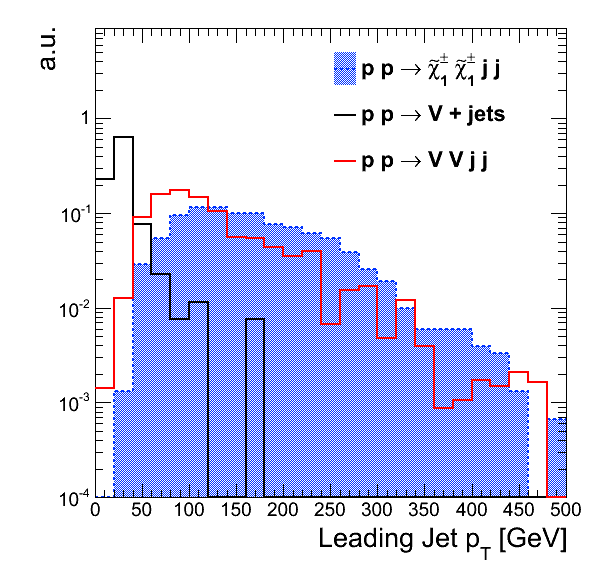}
\caption{$\pT$ distribution of the leading jet normalized to arbitrary units for $\chpm{1} \chpm{1}$ pair production by VBF processes, $V+$jets background, and $VV$ background produced by VBF processes.}
\label{VBFdistributionpTjet}
\end{figure}

In Figures \ref{VBFdistributionMj1j2} and \ref{VBFdistributionpTjet}, the distributions of $M_{j_1j_2}$ and the $\pT$ of the leading jet are displayed for $\chpm{1} \chpm{1}$ pair production by VBF processes, $V+$jets background, and $VV$ background produced by VBF processes. The VBF production channels have large $M_{j_1j_2}$ due to the highly energetic forward jets. The leading jet has high $\pT$ required to create a pair of heavy supersymmetric states. The VBF selections are thus very effective in reducing the background.

Although the central jet veto has been used in the past for VBF Higgs, we
do not employ a central veto
cut in our case as our backgrounds are already small and such a veto in
high pileup conditions can degrade
the signal acceptance and requires extensive study in the future.

We reviewed a successful CMS analysis of forward/backward jets along with
$Z$ production at 7 TeV~\cite{CMS-PAS-FSQ-12-019}, which demonstrated the powerful and efficient identification of
forward/backward jets. Furthermore,
we do not foresee any problems carrying out any CMS analysis with 8 TeV.
For example, $\met$ is calculated
using the entire detector (calorimeter coverage up to $|\eta| = 5$). The same
conclusion holds for ATLAS~\cite{atlaspaper}. Thus,
we expect that both ATLAS and CMS will be able to carry out these VBF analyses
at 8 TeV.


With our proposed  VBF cuts, it is fruitful to divide the study into the $2 \, \tau \, + \met$ and $2 \, l \, + \met$ final states separately.

\subsection{$\geq 2 j \, + \, 2 \tau \, + \, \met$}

For this final state, the following selections are employed in addition to the VBF cuts described above:

$(i)$ Two $\tau$'s with $\pT \geq 20$ GeV in $|\eta| < 2.1$, with $\Delta R (\tau, \tau) > 0.3$. All $\tau$'s considered in this paper are hadronic. The $\tau$ ID efficiency is assumed to be $55\%$ and the jet $\to \tau$ misidentification rate is taken to be $1\%$, both flat over $\pT$ \cite{cmstau}. 
A branching ratio of $100\%$ of $\chpm{1}$ and $\neu{2}$ to $\tilde{\tau}_1$ has been assumed. In realistic models, this branching ratio can be close to  $100\%$.


$(ii)$ $\met > 75$ GeV. This cut is expected to be effective, due to the fact that the main source of $\met$ for signal is the $\neu{1}$ LSP which leaves the detector, while for the background it is the neutrinos produced by leptonic decays of the vector bosons. 

$(iii)$ We also apply a loose $b$-veto which is useful in reducing the $t\overline{t}$ background.

\begin{figure}[!htp]
\centering
\includegraphics[width=3.5in]{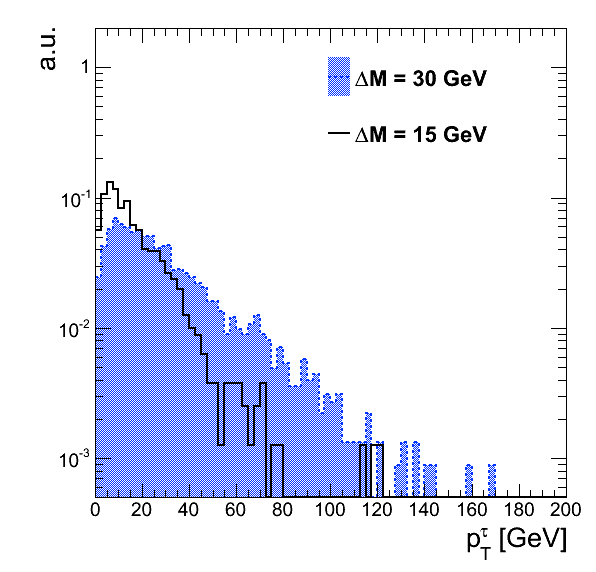}
\caption{$\pT$ of $\tau$ distribution normalized to arbitrary units in $\geq 2j + 2\tau$ final state for $ \Delta M = m_{\tilde{\chi}^{\pm}_1}-m_{\tilde{\chi}^0_1} = 30$ GeV and $15$ GeV.}
\label{VBFTauPt}
\end{figure}


\begin{figure}[!htp]
\centering
\includegraphics[width=3.5in]{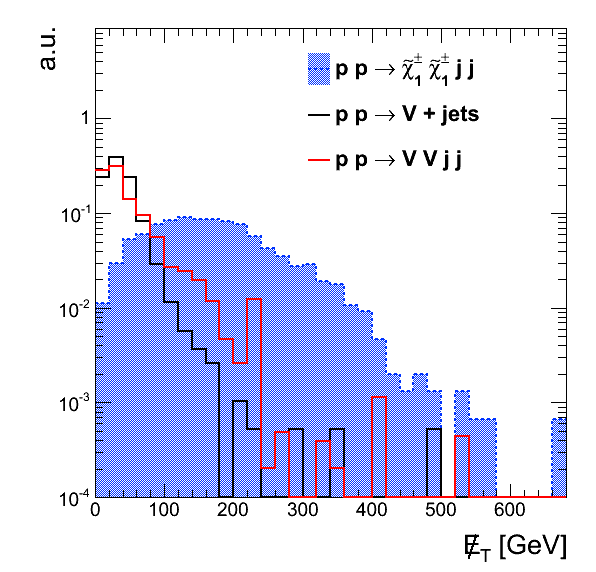}
\caption{$\met$ distribution normalized to arbitrary units for $\chpm{1} \chpm{1}$ pair production by VBF processes, $V+$jets background, and $VV$ background produced by VBF processes.}
\label{VBFdistributionMET}
\end{figure}

In Figure \ref{VBFTauPt}, the normalized distribution of the $\pT$ of $\tau$ is displayed for $ \Delta M = m_{\tilde{\chi}^{\pm}_1}-m_{\tilde{\chi}^0_1} = 30$ GeV and $15$ GeV. For smaller $\Delta M$, the distribution peaks at lower $\pT$ and the signal acceptance is less efficient.



In Figure \ref{VBFdistributionMET}, the $\met$ distribution is displayed for $\chpm{1} \chpm{1}$ pair production by VBF processes, $V+$jets background, and $VV$ background produced by VBF processes. These distributions show that requiring $\met \, > \, 75$ GeV is effective in reducing the background. 


A summary of the effective cross-section at each selection stage of the study is presented in Table \ref{tabletautau} for the main sources of background and inclusive $\chpm{1} \chpm{1}, \chp{1} \chm{1}, \chpm{1}\neu{2}$ and $\neu{2} \neu{2}$ pair production by VBF processes. The VBF and $\met$ cuts are very effective in reducing the background. The significance (at $25$ fb$^{-1}$) for inclusive, opposite-sign, and like-sign $\tau$ pairs are $2.4,\, 1.8,$ and $1.7$, respectively at the benchmark point, where $\Delta M \, = \, m_{\tilde{\tau}_1} - m_{\neu{1}} \, = \, 30$ GeV. For $\Delta M \, = \, 15$ GeV, the significance (at $25$ fb$^{-1}$) for inclusive, opposite-sign, and like-sign $\tau$ pairs are $0.9,\,  0.7,$ and $0.6$, respectively.

We have also considered the $t{\overline{t}}$ background (cross-section $225$ fb) and found that the cross-section for $\geq 2j + 2\tau$ final state is $0.002$ fb (the VBF cut efficiency is $ 10^{-3}$, and the combined efficiency of $\met$, $2 \tau$ inclusive selection and  loose $b$-veto is $\sim 10^{-5}$) which is small compared to the other backgrounds. The VBF cuts are very effective in reducing the background in this case unlike the VBF production of Higgs ~\cite{cms1,atlas1} where much lower jet $\pT$  ($\pT>25$ GeV for ATLAS, $\pT > 30$ GeV for CMS) is used for VBF selection. Also, $M_{j_1j_2}$ is much smaller in the VBF production of Higgs (for example  CMS used $M_{j_1j_2} > 450$ GeV). In the SUSY case,  due to the requirement of larger energy  in the VBF system to produce chargino/neutralino pair, the jet $\pT$ is higher which is utilized to select the signal events~\cite{Choudhury:2003hq}. 

\begin{figure}[!htp]
\centering
\includegraphics[width=3.5in]{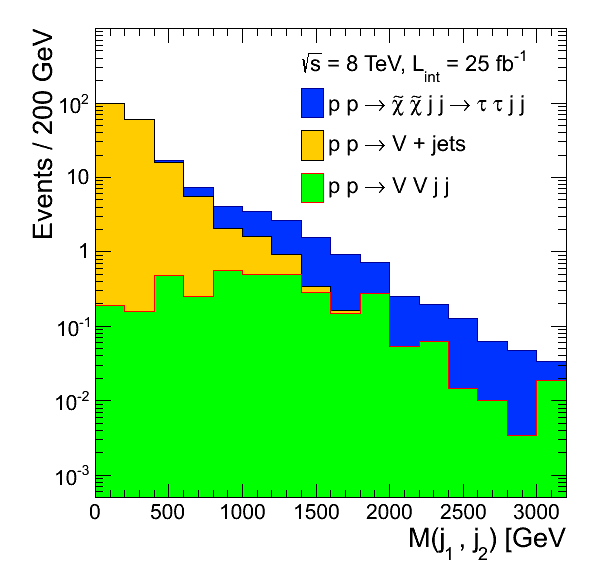}
\caption{Distribution of $M_{j_1j_2}$ for signal and background at $\sqrt{s} = 8$ TeV for $25$ fb$^{-1}$ of integrated luminosity in the $\geq 2j+ 2\tau$ (inclusive) final state. The blue (dark) region displays $M_{j_1j_2}$ distribution for inclusive $\chpm{1} \chpm{1}, \chp{1} \chm{1}, \chp{1}\neu{2}$ and $\neu{2} \neu{2}$ pair production by VBF processes. The yellow (lightest) region displays $V+$jets background. The green (light) region displays $VV$ background produced by VBF processes.}
\label{VBFGoldentautau}
\end{figure}

Figure \ref{VBFGoldentautau} shows the distribution of $M_{j_1j_2}$ for inclusive $\chpm{1} \chpm{1}, \chp{1} \chm{1}, \chp{1}\neu{2}$ and $\neu{2} \neu{2}$ pair production by VBF processes and the main sources of background at $\sqrt{s} = 8$ TeV for $25$ fb$^{-1}$ of integrated luminosity. There is an excess of signal over background above $M_{j_1j_2} \sim 650$ GeV.

\begin{table}[!htp] 
\caption{Summary of the effective cross section (fb) for inclusive $\chpm{1} \chpm{1}, \chp{1} \chm{1}, \chp{1}\neu{2}$ and $\neu{2} \neu{2}$ pair production by VBF processes and main backgrounds in the $\geq 2j + 2\tau$ final state at $\sqrt{s} = 8$ TeV. Results for same-sign and oppositely-signed final state $\tau$ pair, as well as inclusive study, are shown. The final $t \overline{t}$ cross-section after all cuts is $0.002$ fb in the $\geq 2j + 2\tau$ (inclusive) final state at $\sqrt{s} = 8$ TeV, and this has been neglected in the computation of the significance. Masses and momenta are in GeV. The significance is displayed for $25$ fb$^{-1}$.}
\label{tabletautau}
\begin{center}
\begin{tabular}{c c c c c c} 
\hline \hline 
                    

                               &Signal    &\,\, $Z+$jets &\,\, $W+$jets &\,\, $WW$ &\,\, $WZ$ \\
          
\hline 
              \hline  \\
                    



                     VBF cuts            &$4.61$   & $10.9$                        &$3.70 \times 10^3$                      &$97.0$  &$19.0$      \\      
      $\met > 75, b$-veto                 &$4.33$   & $0.27$                        &$5.29 \times 10^2$                      &$17.6$  &$3.45$      \\
      \hline
       $2 \, \tau,$ inclusive                  & $0.45$       & $0.06$                             &$0.23$                           &$0.09$       &$0.04$          \\   
        $(S/\sqrt{S+B})$                       & \multicolumn{5}{c}{$2.4$} \\ 
       \hline
      $\tau^{\pm} \tau^{\pm}$                  &$0.21$       &$0$                             &$0.11$                           &$0.02$       &$0.01$          \\                          
    $(S/\sqrt{S+B})$                       & \multicolumn{5}{c}{$1.8$} \\         
\hline
      $\tau^{\pm} \tau^{\mp}$                  &$0.24$       &$0.06$                             &$0.12$                           &$0.07$       &$0.03$          \\   
     $(S/\sqrt{S+B})$                       & \multicolumn{5}{c}{$1.7$} \\       
\hline \hline

\end{tabular}
\end{center}
\end{table}



\subsection{$\geq 2 j \, + \, 2 \mu \, + \, \met$}

In the previous section, the VBF search strategy in the multi-$\tau$ final state has been outlined. Although the search with $\tau$ leptons are expected to be the most daunting case amongst the three generation of leptons, it is worthwhile to study the potential sensitivity in cases where the charginos and neutralinos preferentially decay to the first two generation of sleptons.

In this subsection, results for event selection with $\geq 2 j \, + \, 2 \mu \, + \, \met$ are presented. A branching ratio of $100\%$ of $\chpm{1}$ and $\neu{2}$ to $\tilde{\mu}$ has been assumed. After VBF selections, the following selections are employed:

$(i)$ Two isolated $\mu$'s with $\pT \geq 20$ GeV and $\pT \geq 15$ GeV in $|\eta| < 2.1$.

$(ii)$ $\met > 75$ GeV.  

$(iii)$ We also apply a loose $b$-veto which is useful in reducing the $t\overline{t}$ background.

A summary of the effective cross-section at each selection stage of the study is presented in Table \ref{tablemumu} for the main sources of background and inclusive $\chpm{1} \chpm{1}, \chp{1} \chm{1}, \chp{1}\neu{2}$ and $\neu{2} \neu{2}$ pair production by VBF processes. The VBF and $\met$ cuts are very effective in reducing the background. 
The $t\overline{t}$ background  cross-section for $\geq 2j + 2\mu$ final state is $0.04$ fb (the VBF cut efficiency is $10^{-3}$, and the combined efficiency of $\met$ cut, 2$\mu$ inclusive event selection, and loose $b$-veto is $\sim 2\times 10^{-4}$) which is small compared to the other backgrounds. Also we note that, the inclusive and opposite sign muon cross-sections are same, 0.04 fb, but the same sign dimuon cross-section is almost zero.

The significances (at $25$ fb$^{-1}$) for inclusive, opposite sign, and like-sign $\mu$ pairs are $6.0, \, 4.4,$ and $4.1$, respectively. For $m_{\tilde{\chi}^{\pm}_1} \, \sim \, m_{\tilde{\chi}^0_2} \, \sim 260$ GeV, the significances (given by $S/\sqrt{S+B}$) for inclusive, opposite-sign, and like-sign $\mu$ pairs are approximately $2.7, \, 1.8,$ and $1.7$, respectively (at $25$ fb$^{-1}$).

\begin{figure}[!htp]
\centering
\includegraphics[width=3.5in]{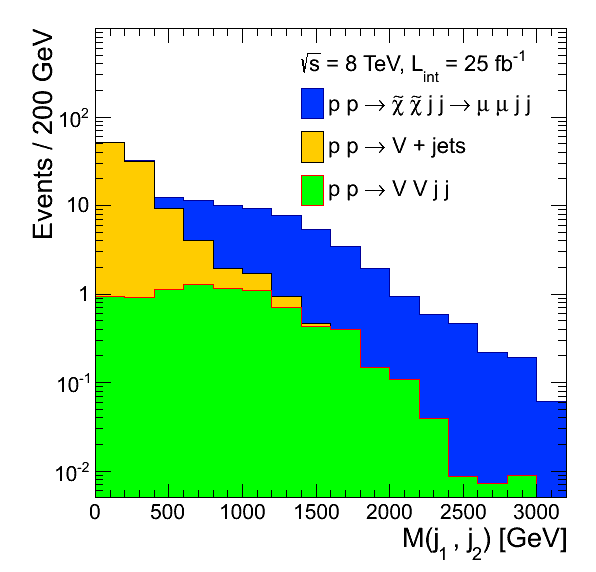}
\caption{Distribution of $M_{j_1j_2}$ for signal and background at $\sqrt{s} = 8$ TeV for $25$ fb$^{-1}$ of integrated luminosity in the $\geq 2j+ 2\mu$ (inclusive) final state. The blue (dark) region displays $M_{j_1j_2}$ distribution for inclusive $\chpm{1} \chpm{1}, \chp{1} \chm{1}, \chp{1}\neu{2}$ and $\neu{2} \neu{2}$ pair production by VBF processes. The yellow (lightest) region displays $V+$ jets background. The green (light) region displays $VV$ background produced by VBF processes.}
\label{VBFGoldenmumu}
\end{figure}

Figure \ref{VBFGoldenmumu} shows the distribution of $M_{j_1j_2}$ for inclusive $\chpm{1} \chpm{1}, \chp{1} \chm{1}, \chp{1}\neu{2}$ and $\neu{2} \neu{2}$ pair production by VBF processes and main sources of background at $\sqrt{s} = 8$ TeV for $25$ fb$^{-1}$ of integrated luminosity. There is an excess of signal over background above $M_{j_1j_2} \sim 650$ GeV.

\begin{table}[!htp] 
\caption{Summary of the effective cross section (fb) for inclusive $\chpm{1} \chpm{1}, \chp{1} \chm{1}, \chp{1}\neu{2}$ and $\neu{2} \neu{2}$ pair production by VBF processes and main backgrounds in the $\geq 2j + 2\mu$ final state at $\sqrt{s} = 8$ TeV. Results for same-sign and oppositely-signed final state $\mu$ pair, as well as inclusive study, are shown. The final $t \overline{t}$ cross-section after all cuts is $0.04$ fb in the $\geq 2j + 2\mu$ (inclusive) final state at $\sqrt{s} = 8$ TeV, and this has been added in the computation of the significance (inclusive and opposite sign muon cross-sections are same, 0.04 fb, but the same sign dimuon cross-section is almost zero). Masses and momenta are in GeV.  The significance is displayed for $25$ fb$^{-1}$.}
\label{tablemumu}
\begin{center}
\begin{tabular}{c c c c c c} 
\hline \hline 
                    

                               &Signal    &\,\, $Z+$jets &\,\, $W+$jets &\,\, $WW$ &\,\, $WZ$ \\
          
\hline 
              \hline  \\

                     VBF cuts            &$4.61$   & $10.9$                        &$3.70 \times 10^3$                      &$97.0$  &$19.0$      \\      
      $\met > 75$                 &$4.33$   & $0.27$                        &$5.29 \times 10^2$                      &$17.6$  &$3.45$      \\
      \hline
       $2 \, \mu,$ inclusive                  & $1.83$       & $0.15$                             &$0$                           &$0.12$       &$0.19$          \\   
        $(S/\sqrt{S+B})$                       & \multicolumn{5}{c}{$6.0$} \\ 
       \hline
      $\mu^{\pm} \mu^{\pm}$                  &$0.87$       &$0$                             &$0$                           &$0.03$       &$0.05$          \\                          
    $(S/\sqrt{S+B})$                       & \multicolumn{5}{c}{$4.5$} \\         
\hline
      $\mu^{\pm} \mu^{\mp}$                  &$0.96$       &$0.15$                             &$0$                           &$0.09$       &$0.14$          \\   
     $(S/\sqrt{S+B})$                       & \multicolumn{5}{c}{$4.1$} \\       
\hline \hline

\end{tabular}
\end{center}
\end{table}

\section{Conclusion} \label{conclusion}

This paper has investigated the production of $\chpm{1}$ and $\neu{2}$ by VBF processes at the LHC at $\sqrt{s} = 8$ TeV. The presence of high $E_T$ forward jets in opposite hemispheres with large dijet invariant mass is used to identify the VBF production. Kinematic requirements to search for signals of these supersymmetric particles above SM background arising from VBF and non-VBF processes have been developed. They have been shown to be effective in searching for $\chpm{1}$ and $\neu{2}$ in both $2l$ as well as $2 \tau$ final states. 

Searches for the EW sector in $\tau$ final states in Drell-Yan production face the challenge of controlling the level of backgrounds due to the larger $\tau$ misidentification rate as well as maintaining low enough $p_T$ thresholds for triggering. The VBF searches  are capable of reducing the background to manageable levels and thus probing multi-$\tau$ final states.

At the benchmark point ($m_{\tilde{\chi}^{\pm}_1} \, \sim \, m_{\tilde{\chi}^0_2} \, = 181$ GeV, $m_{\tilde{\tau}_1} = 130$ GeV, and $m_{\tilde{\chi}^0_1} = 100$ GeV), the significances (at $25$ fb$^{-1}$) for inclusive, opposite-sign, and like-sign $\tau$ pairs are $2.4,\, 1.8,$ and $1.7$, respectively, for $\Delta M \, = \, m_{\tilde{\tau}_1} - m_{\neu{1}} \, = \, 30$ GeV, while 0.9, 0.7 and 0.6, respectively, for $\Delta M \, = \, 15$ GeV.

The significances for inclusive, opposite-sign, and like-sign $\mu$ pairs are $6.0, \, 4.5,$ and $4.1$, respectively, at the benchmark point. 
However,  the significances   are reduced to $2.7, \, 1.8,$ and $1.7$, respectively, for $m_{\tilde{\chi}^{\pm}_1} \, \sim \, m_{\tilde{\chi}^0_2} \, \sim 260$ GeV.

The next-to-leading order QCD corrections to the VBF electroweak
production cross sections
have not been considered. The inclusion of the K factor, which is very
modest for VBF production
($\sim$ 5\%)~[29], would improve the signal significance~\cite{kfactor}.


With increasing instantaneous luminosity, both ATLAS and CMS experiments are raising their $p_T$  thresholds for triggering objects. The VBF trigger offers a promising route to probe supersymmetric production free from trigger bias. This is complementary to the existing LHC searches based on Drell-Yan production.

\section{Acknowledgements}

We would like to thank Sean Wu for useful discussions. This work is supported in part by the DOE grant DE-FG02-95ER40917, by the World Class University (WCU) project through the National Research Foundation (NRF) of Korea funded by the Ministry of Education, Science \& Technology (grant No. R32-2008-000-20001-0), and by the National Science Foundation grant  PHY-1206044.


\end{document}